# Evidences of Mott physics in iron pnictides from x-ray spectroscopy

S. Lafuerza[1,*], H. Gretarsson[2], F. Hardy[4], T. Wolf[4], C. Meingast[4], G. Giovannetti[5], M. Capone[5], A. S. Sefat[6] Y. –J. Kim[3], P. Glatzel[1,*] and L. de' Medici[1,*]

[1]*ESRF – The European Synchrotron, CS40220, F-38043 Grenoble Cedex 9, France*
[2]*Max-Planck-Institut für Festkörperforschung, Heisenbergstraße 1, D-70569 Stuttgart, Germany*
[3]*Department of Physics, University of Toronto, 60 St. George St., Toronto, Ontario, M5S 1A7, Canada*
[4]*Karlsruher Institut für Technologie, Institut für Festkörperphysik, 76021 Karlsruhe, Germany*
[5]*CNR-IOM-Democritos National Simulation Centre and International School for Advanced Studies (SISSA), Via Bonomea 265, I-34136, Trieste, Italy*
[6]*Materials Science and Technology Division, Oak Ridge National Laboratory, Oak Ridge, Tennessee 37831-6114, USA*
*E-mail: sara.lafuerza@esrf.fr, demedici@esrf.fr, pieter.glatzel@esrf.fr

**Abstract**

X-ray emission and absorption spectroscopies are jointly used as fast probes to determine the evolution as a function of doping of the fluctuating local magnetic moments in K- and Cr- hole-doped $BaFe_2As_2$. An increase in the local moment with hole-doping is found, supporting the theoretical scenario in which a Mott insulating state that would be realized for half-filled conduction bands has an influence throughout the phase diagram of these iron-pnictides.

A scenario has recently been proposed in the field of Fe-based superconductors, where the low energy physics is described in terms of a coexistence of strongly and weakly correlated electrons [1]. Such a picture provides a framework to understand a wealth of experiments on quasiparticle mass enhancements (specific heat, low-frequency optical conductivity, angle-resolved photoemission spectroscopy, quantum oscillations etc.) in electron- and hole-doped $BaFe_2As_2$ [1]. At the center of this model is a Mott-insulating state favoured by Hund's coupling that would be realized if the system could be doped until the $d^5$ configuration (half-filling of the conduction bands, which have predominant character of the 5 Fe-3*d* orbitals). In this scenario the Mott insulator extends its influence in the actual range of doping in which these materials exist, from $d^{5.5}$ (for hole-doped compounds, like $KFe_2As_2$) to $d^6$ (for the stoichiometric parent compound $BaFe_2As_2$) and beyond (for electron-doped compounds). In doped $BaFe_2As_2$ quasiparticle masses, while differing depending on the orbital character, overall increase continuously from the electron-doped to the hole-doped side of the phase diagram [2]. Besides enhancing quasiparticle masses, Mott-induced correlations (particularly when Hund's coupling is strong) are expected to enhance the local, fluctuating moments in the paramagnetic phase. The Hund's coupling-driven Mott insulator with 5 electrons in 5 orbitals would be in the high-spin state ($S$ = 5/2) [3,4]. The neighbouring metallic phase is expected to be dominated by high-spin configurations, with thus a magnetic moment building up, with hole-doping, towards this saturated value. An experimental confirmation of this high-energy counterpart of the aforementioned quasiparticle behaviour would highlight the role of Mott physics in these materials.

A direct method to probe the local magnetic moment in transition metal compounds is $K\beta$ core-to-core (CTC) x-ray emission spectroscopy (XES) that is sensitive to the local 3*d* spin because of the intra-atomic 3*p*-3*d* exchange interaction [5,6]. This gives the possibility to probe the magnetic moment even in the paramagnetic state where the local moments are fluctuating. We have applied this technique to investigate whether the Fe local magnetic moment in the paramagnetic phase raises with increasing hole-doping, as predicted by the Hund-Mott physics scenario [1,3,7].

Two different families of hole-doped $BaFe_2As_2$ compounds have been studied: $Ba_{1-x}K_xFe_2As_2$ (out-of-plane doping) and $Ba(Fe_{1-x}Cr_x)_2As_2$ (in-plane doping). At room temperature, the parent compound $BaFe_2As_2$ crystallizes in a tetragonal structure that consists of $(FeAs)^-$ iron arsenide layers separated by $Ba^{2+}$ ions. In $Ba_{1-x}K_xFe_2As_2$, partial substitution of $Ba^{2+}$ with $K^+$ introduces holes in the $(FeAs)^-$ layers and superconductivity emerges [8,9]. In contrast, the $Ba(Fe_{1-x}Cr_x)_2As_2$ compounds, where in-plane Cr substitution also induces hole-doping, are not superconducting [10]. $K^+$ injects one electron less than $Ba^{2+}$ (the number of doped holes/Fe scale with $x/2$) while $Cr^{2+}$ ($3d^4$) has two $3d$ electrons less than $Fe^{2+}$ ($3d^6$) (the number of doped holes/Fe scale with $2x$).

The unoccupied density of states around Fe can be probed by means of *K* edge x-ray absorption spectroscopy (XAS) providing information that is complementary to $K\beta$ spectroscopy. We have recorded high-energy resolution fluorescence detected XAS on all K- and Cr-doped samples.

All samples were plate-like single crystals grown by the self-flux technique as detailed elsewhere [10–12]. Two sets of $Ba_{1-x}K_xFe_2As_2$ were measured covering the K-doping range $0 \leq x \leq 1$ (set 1 [11] with x = 0, 0.18, 0.6, 0.85, 1; and set 2 [12] with x = 0.08, 0.16, 0.25, 0.37) and one set of $Ba(Fe_{1-x}Cr_x)_2As_2$ with x = 0.026, 0.07, 0.12, 0.2, 0.3, 0.47. In addition, we have measured a single crystal of FeCrAs where Fe is non-magnetic [13–15]. The experiment was conducted at beamline ID26 of the European Synchrotron Radiation Facility (ESRF). The incident energy was tuned through the Fe *K* edge (7112 eV) by means of a pair of cryogenically cooled Si(111) monochromator crystals. The inelastically scattered photons were analyzed using a set of five spherically bent Ge(620) crystals that were arranged with the sample and detector (avalanche photodiode) in a vertical Rowland geometry ($R$ = 1 m). The surface of the samples was aligned to 45º with respect to the incident beam direction, i.e. the angle between the electric field of the linearly polarized x-rays and the sample [001] direction. Non-resonant Fe $K\beta$ XES measurements were recorded at incident energy 7200 eV. High-energy resolution fluorescence-detected x-ray absorption near edge structure (HERFD-XANES) spectra were obtained by setting the emission energy to the maximum of the $K\beta_{1,3}$ line while scanning the incoming energy through the absorption edge. The resulting spectrum is an approximation to the 1*s* photoabsorption cross section with sharper spectral features than in standard XAS [5,6]. The total experimental broadening, determined as full width at half maximum of the elastic profiles, was 1.3 eV. All the measurements were performed at room temperature. Both XES and HERFD-XANES spectra have been normalized in area using the ranges 7025-7080 eV and 7105-7190 eV respectively.

Non-resonant Fe $K\beta$ CTC XES spectra have been measured as a function of doping in $Ba_{1-x}K_xFe_2As_2$ ($0 \leq x \leq 1$) and $Ba(Fe_{1-x}Cr_x)_2As_2$ ($0.026 \leq x \leq 0.47$). Figure 1(a) shows the Fe $K\beta$

CTC XES spectra for undoped $BaFe_2As_2$, the highest K- and Cr-content samples as representative examples and the compound FeCrAs. The CTC $K\beta$ XES arise from a $3p \rightarrow 1s$ transition and the spectral features are separated into the strong $K\beta_{1,3}$ peak and a broad $K\beta$' shoulder at lower emitted energy. Compared to FeCrAs, where a non-magnetic Fe is also in tetrahedral coordination with As [13–15], the spectra of $BaFe_2As_2$, $KFe_2As_2$ and $Ba(Fe_{0.53}Cr_{0.47})_2As_2$ show the $K\beta_{1,3}$ peak shifted to higher energies and a more intense $K\beta$' shoulder. This illustrates that in the latter compounds the Fe atoms carry spin magnetic moments. In insulators a small intensity of the $K\beta$' feature is taken as an indication of a low-spin state and this spectral signature has been widely used to identify transitions from high- to low-spin as a function of pressure [16], temperature [17] or composition [18]. For metallic systems it has been observed that the $K\beta$' feature is very broad [19]. The splitting between the $K\beta_{1,3}$ and $K\beta$' is approximately given by $\Delta E = J(2S + 1)$ [20] where $J$ is the exchange integral between the electrons in the $3p$ and $3d$ shells. It becomes visible in the difference spectra of Figure 1 and is about 10-15 eV in our compounds which is compatible with a large $3d$ magnetic moment. The main change in the spectra upon doping is a shift of the $K\beta_{1,3}$ maximum to higher energies (see inset in Figure 1(a)). In order to obtain the quantitative evolution of the Fe local moment as a function of hole-doping from the CTC $K\beta$ XES, we have applied the data reduction technique known as integrated absolute difference (IAD) [21]. This method allows obtaining the relative change in the local spin magnetic moment by integrating the absolute value of the difference between sample spectrum and a reference spectrum. We chose $BaFe_2As_2$ as the reference and our IAD values thus represent the change in the local spin moment between each doped sample of the $Ba_{1-x}K_xFe_2As_2$ and $Ba(Fe_{1-x}Cr_x)_2As_2$ series and the undoped compound. In Figure 1(a) are shown as example the difference spectra for the highest K- and Cr-content samples respectively. The corresponding zero signal in each case (*i.e.* the difference spectra between equivalent measurements for the same sample) was checked to assess the experimental error. In Figure 1(b) the IAD values are plotted as a function of doped holes/Fe, where the error bars are as estimated from comparison of equivalent measurements taken in the same conditions either during a different experimental session or in a sample from a different batch. In both cases a clear increase of the IAD value with hole-doping is observed, although the trend is somewhat smaller for the K-doped samples. If the FeCrAs spectrum is used as the reference for the IAD analysis as done in [22,23], the lineshape of the difference spectra for both the K- and Cr-doped compounds has the typical reported [22–26] shape showing a bump in the energy position of the $K\beta$' satellite and the downward-upward feature around the $K\beta_{1,3}$ peak. Interestingly, when using the parent compound $BaFe_2As_2$ as the reference instead, more structure is revealed. The K-doped compounds show a different modulation with an extra upward peak at the low energy side of the $K\beta_{1,3}$ maximum while the Cr-doped compounds still follow the typical shape (see Fig. 1(a)). A possible explanation is that this modulation for the K-doped samples might be ascribed to differences in the *dd*-interactions between the two types of hole-doping.

Figure 2 shows the experimental HERFD-XANES spectra for the $Ba_{1-x}K_xFe_2As_2$ and $Ba(Fe_{1-x}Cr_x)_2As_2$ series. The spectrum of $BaFe_2As_2$ (x = 0) shows six distinctive features denoted as *A, B, C, D, E* and *F* in the figure that are in agreement with previously reported conventional XANES spectra for the same compound [22,23]. All these features exhibit significant changes in the K-doped samples. In particular, the *A* peak shows an increase accompanied by a small shift to higher energy (of about 0. 2 eV for x = 1). This result was also reported in standard XANES spectra for the $0 \leq x \leq 0.5$ doping range [27]. Upon Cr-doping the only significant variation is an intensity decrease in the *A* peak while the observed peak shift is much smaller ($\leq$ 0.05 eV). The *A* peak intensity as a function of doped holes/Fe is shown in the inset of Figure 2. This feature just below the main rising edge occurs in the energy range of Fe 1*s* to 3*d* transitions. However, if inversion symmetry is absent as it is the case in the $FeAs_4$ tetrahedra, the metal 3*d* and 4*p* orbitals can mix resulting in strong $1s \rightarrow 4p$ dipolar transitions that dominate over the quadrupolar contribution at the same energy. Another dipole contribution to the pre-edge may arise from ligand mediated non-local transitions which result from the mixing of the metal (M) 4*p* character into the 3*d* orbitals of a neighbouring metal atom (M') [29]. The metal 4*p*-3*d* *intrasite* (M-M) or *intersite* (M-M') orbital mixing result in a large dipole contribution that originates from transitions to unoccupied metal *d* – ligand *p* hybrid bands. Accordingly, previous works in related Fe based superconductors ascribed the *A* peak to the Fe-ligand orbital mixing [23–26,30,31]. In the supplemental material we include calculations of our HERFD-XANES spectra showing that the *A* peak has mostly dipole character and that it arises when the four As nearest neighbours conforming the $FeAs_4$ tetrahedra are included in the structure, which corroborates its origin in the Fe 3*d* – As 4*p* mixing. The simulations nicely reproduce the behaviour of the experimental spectra upon K- and Cr-doping. Our results suggest an enhancement of the Fe 3*d* – As 4*p* orbital mixing with K-doping and a modest decrease with Cr-doping. We have investigated whether the behaviour of the *A* peak is correlated with the changes in the in-plane *a* lattice parameter of the tetragonal structure. In the inset of Figure 2 (right axis) we present the variation of the lattice parameter as a function of doped holes/Fe. The data have been taken from diffraction experimental results reported in refs. [9] and [10] for $Ba_{1-x}K_xFe_2As_2$ and $Ba(Fe_{1-x}Cr_x)_2As_2$ respectively. $Ba^{2+}$ substitution by $K^+$ out-of-plane induces a shrinking of the in-plane *a* lattice parameter along with the shortening of the Fe-As and Fe-Fe interatomic distances, whereas $Fe^{2+}$ replacement by $Cr^{2+}$ barely induces any structural changes in the Fe-As plane. Clearly, the *A* peak intensity evolution is strongly correlated with the changes in the in-plane lattice parameter. In $Ba_{1-x}K_xFe_2As_2$, as a result of the shrinkage of the lattice parameter and with that of the interatomic distances, the Fe 3*d* – As 4*p* orbital mixing increases as reflected in the rise of the *A* peak intensity. Compared to K-doping, Cr-doping hardly affects the lattice parameter except for a subtle increment that can account for a small decrease of the orbital mixing and in turn of the *A* peak intensity. We include in the supplementary material the Fe $K\beta$ valence-to-core (VTC) XES measurements that also indicate changes in the occupied density of states correlated with the evolution of the lattice parameter and the covalency variations.

A finite IAD value is generally attributed to a change in the spin magnetic moment as it reflects a modification of the $K\beta'$-$K\beta_{1,3}$ splitting [20]. This splitting depends on the 3*p* – 3*d* exchange integral *J* and the valence shell spin. Stronger metal-ligand covalency may be interpreted in terms of a larger radial distribution of the effective (Wannier) metal *d*-orbitals used in the description of the electron density, implying a decrease of *J* which translates into a reduction of the IAD value. Indeed the influence of the metal – ligand covalency on the $K\beta$ CTC spectra has been reported earlier in different works [5,32,33]. Therefore, while *J* can be assumed to vary very little with Cr-doping, it decreases along the $Ba_{1-x}K_xFe_2As_2$ series with increasing K-content as a result of the spread of the Wannier orbitals of Fe 3*d* character that also manifests itself in the increase of the 3*d*–4*p* orbital mixing. Since the intensity of the *A* peak in the HERFD-XANES spectra reflects the amount of 3*d*-4*p* mixing and implicitly the variations in the 3*p* – 3*d* exchange integral, we have performed a rescaling of the IAD multiplying by the intensity of this peak for each sample $I_{A(x)}$. For the sake of comparison to the undoped compound, we have normalized to $I_{A(x=0)}$. The rescaled IAD values are shown in Figure 3. In this way we get a unified scenario between the two types of hole-doping, which is globally in agreement with the increase in the local magnetic moment predicted by the theoretical scenario based on the occurrence of a Hund-promoted Mott insulator at half-filling [1,3,13]. This increase can be qualitatively understood by comparing to a theoretical modelling of the local magnetic moment obtained within Density Functional Theory (DFT) combined with Slave-spin mean-field (see supplementary material for details on the method). Indeed in the inset of Fig. 3 is reported, as a function of the interaction strength *U*, the calculated spin-spin local correlation function $\langle S_z S_z \rangle$, which increases monotonically with the total local magnetic moment (e.g. at low *U*, $S(S+1) = 3\langle SzSz \rangle$). It is clear that in the uncorrelated ($U = 0$) limit (as described by DFT) the change in local magnetic moment between the compounds is negligible. On the contrary, at large *U* a different behaviour is found, with the moment increasing towards saturated values, signalling the zone influenced by the Mott insulator realized at half-filling. In this strongly correlated zone the magnetic moment increases steadily with hole-doping, thus the saturated value for $KFe_2As_2$ is larger than for $BaFe_2As_2$. This effect is due to the hole-doping, while the change in the lattice structure between the two compounds has a minor influence in the opposite direction. This is visible as a slightly smaller value of the magnetic moment for $KFe_2As_2$ at small to intermediate *U*, caused by a small increase in the DFT bare bandwidth of the latter, compared to $BaFe_2As_2$, which embodies the increased covalency. The doping effect is however dominant at larger *U*. The absence of this covalency effect due to the lack of structural changes in the Cr-doped samples thus favours the slightly larger increase of the moment compared to the K-doped case corroborating the trend we observe in the measurements.

The XES-XAS study presented in this Letter confirms the increase of the Fe local moment upon hole-doping in $BaFe_2As_2$, providing experimental evidence of the Hund-Mott physics model [1,3,7] which is an important input for the community studying Fe-based superconductors. Besides, by combining results from x-ray emission with x-ray absorption spectroscopy, we have been able to disentangle the different influence of structural and correlation effects on the local spin moment in K- and Cr-doped $BaFe_2As_2$.

The authors thank ESRF for granting beam time and ID26 beamline staff for their assistance during the experiments. This research was supported in part by the National Science Foundation under Grant No. NSF PHY11-25915, through the support of LdM for the participation to the KITP IRONICS14 workshop. LdM would also like to thank discussions with M. Casula and K. Gilmore, M.B. Lepetit. GG and MC acknowledge funding by FP7/ERC Grant ``SUPERBAD" (GA 240524) and SISSA/CNR project “Superconductivity, Ferroelectricity and Magnetism in bad metals”. YJK acknowledges the support from Natural Science and Engineering Research Council of Canada.

## FIGURE CAPTIONS

**Figure 1.** (a) Normalized Fe $K\beta$ CTC XES spectra of $BaFe_2As_2$, $KFe_2As_2$, $Ba(Fe_{0.53}Cr_{0.47})_2As_2$ and FeCrAs together with the difference spectra between $BaFe_2As_2$ and the $KFe_2As_2$ (blue) and $Ba(Fe_{0.53}Cr_{0.47})_2As_2$ (green) compounds respectively, multiplied by a factor of 5. Inset: detail of the $K\beta_{1,3}$ maxima. (b) IAD values derived from the Fe $K\beta$ CTC XES spectra of the $Ba_{1-x}K_xFe_2As_2$ and $Ba(Fe_{1-x}Cr_x)_2As_2$ series normalized to the spectral area, shown as a function of doped holes/Fe (x/2 and 2x for K- and Cr-doping series respectively). Circles and triangles differentiate the K-doped samples from sets 1 and 2 respectively.

**Figure 2.** Normalized HERFD-XANES spectra at the Fe *K* edge of the $Ba_{1-x}K_xFe_2As_2$ ($0 \leq x \leq 1$) and $Ba(Fe_{1-x}Cr_x)_2As_2$ ($0.026 \leq x \leq 0.47$) series. Inset: Evolution of the *A* peak intensity (left axis) as a function of doped holes/Fe (x/2 and 2x for K- and Cr-doping series respectively). Circles and triangles differentiate the K-doped samples from sets 1 and 2 respectively. On the right axis the data of the tetragonal *a* lattice parameter from refs. [9] and [10] is plotted for a direct comparison.

**Figure 3.** Renormalized IAD values taking into account the changes in the *A* peak intensity derived from the HERFD-XANES data. Inset: DFT+Slave-spin mean-field calculations of the local magnetic moment for $BaFe_2As_2$ (Fe in $d^6$ configuration) and $KFe_2As_2$ (Fe in $d^{5.5}$ configuration) as a function of the interaction strength *U*, for Hund's coupling $J/U$=0.25 [1]. A strong increase of the local moment with hole-doping is only observed in the region ($U$>2.5eV) influenced by the Mott insulating phase realized for a half-filled configuration $d^5$. For further details see the supplementary material.

**Figure 1.**

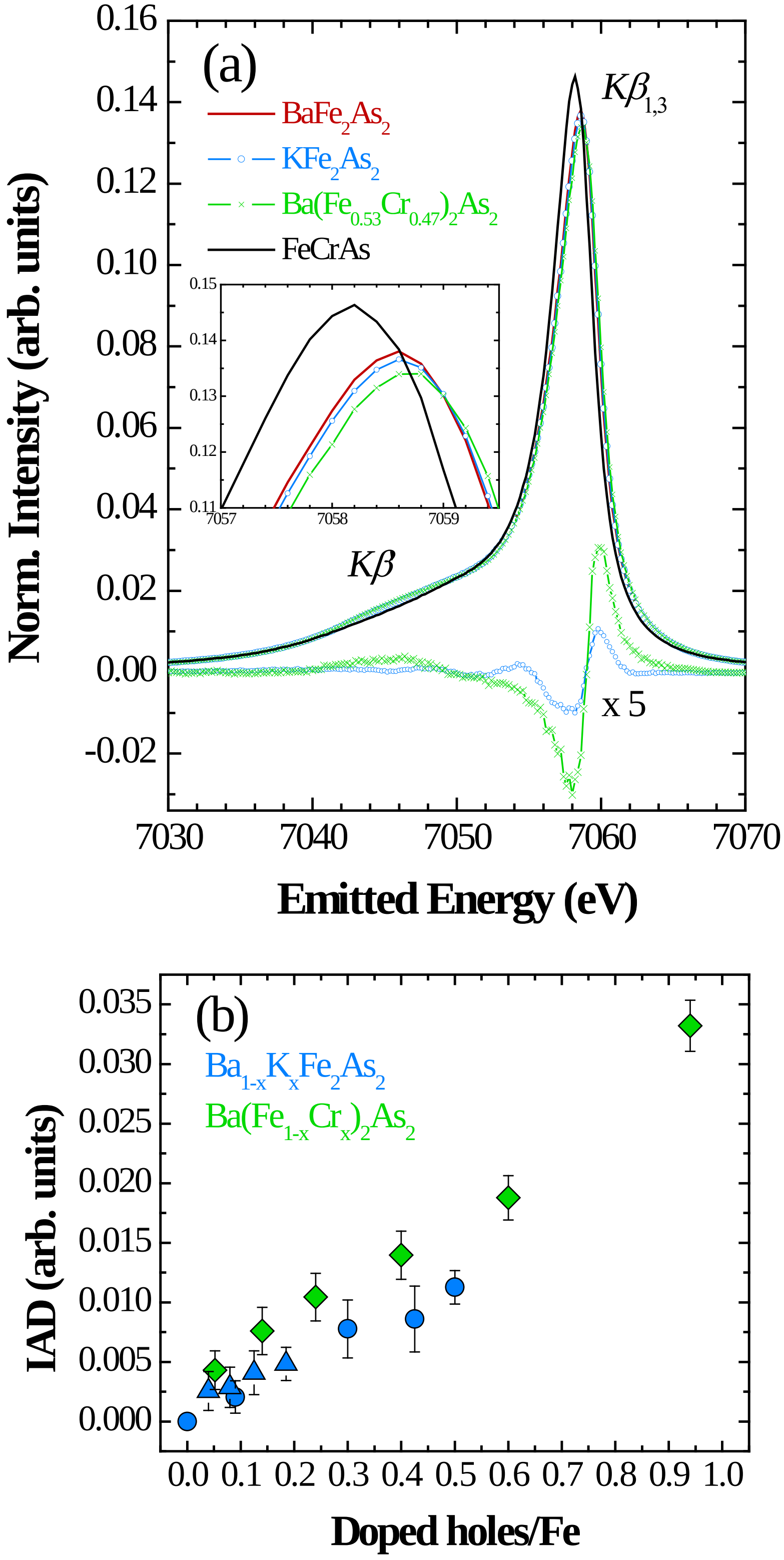

(a)
BaFe2As2
KFe2As2
Ba(Fe0.53Cr0.47)2As2
FeCrAs
Kβ1,3
Kβ'
x 5
Norm. Intensity (arb. units)
Emitted Energy (eV)
0.16
0.14
0.12
0.10
0.08
0.06
0.04
0.02
0.00
-0.02
7030
7040
7050
7060
7070
0.15
0.14
0.13
0.12
0.11
7057
7058
7059
(b)
Ba1-xKxFe2As2
Ba(Fe1-xCrx)2As2
IAD (arb. units)
Doped holes/Fe
0.035
0.030
0.025
0.020
0.015
0.010
0.005
0.000
0.0 0.1 0.2 0.3 0.4 0.5 0.6 0.7 0.8 0.9 1.0

**Figure 2.**

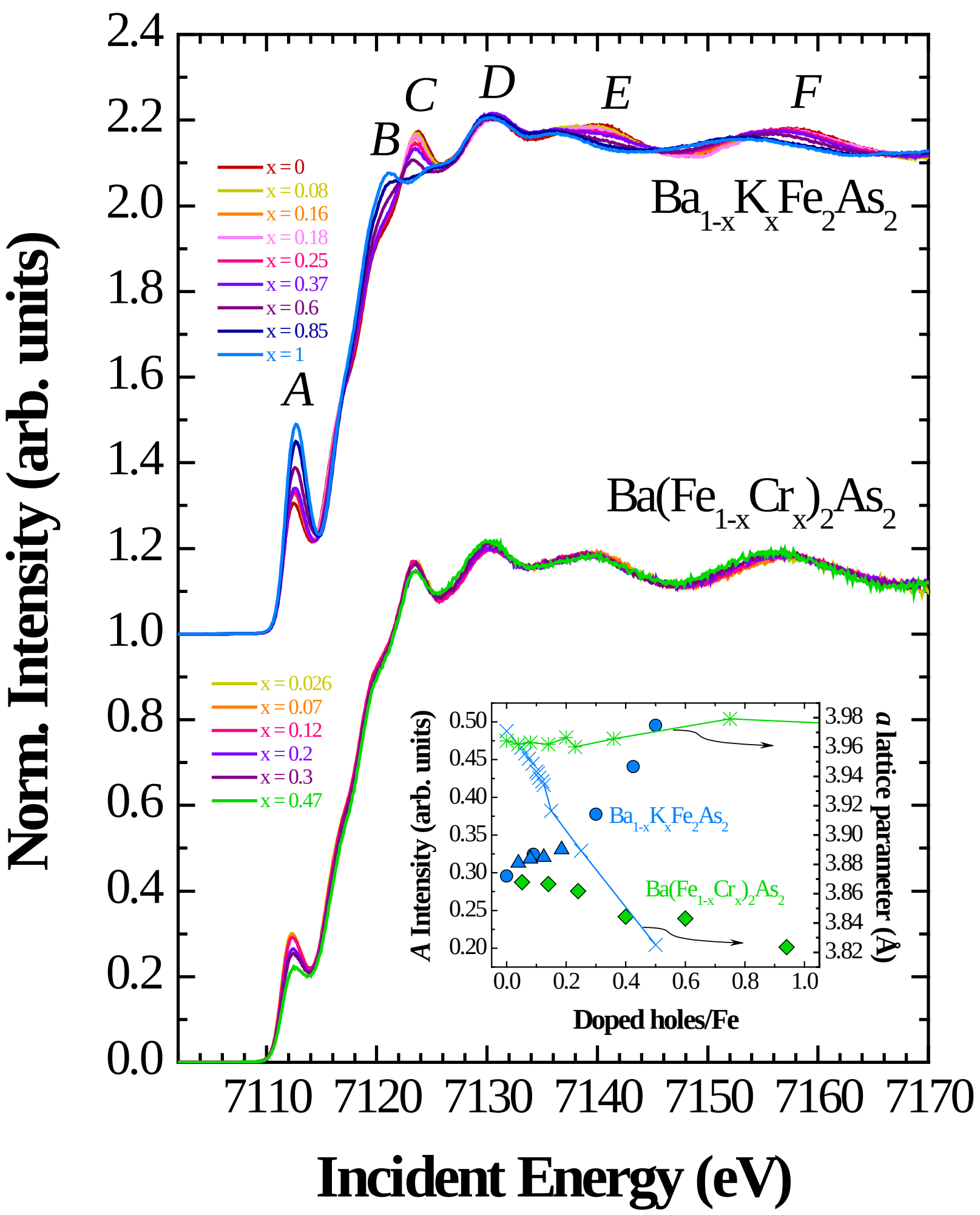

**Figure 3.**

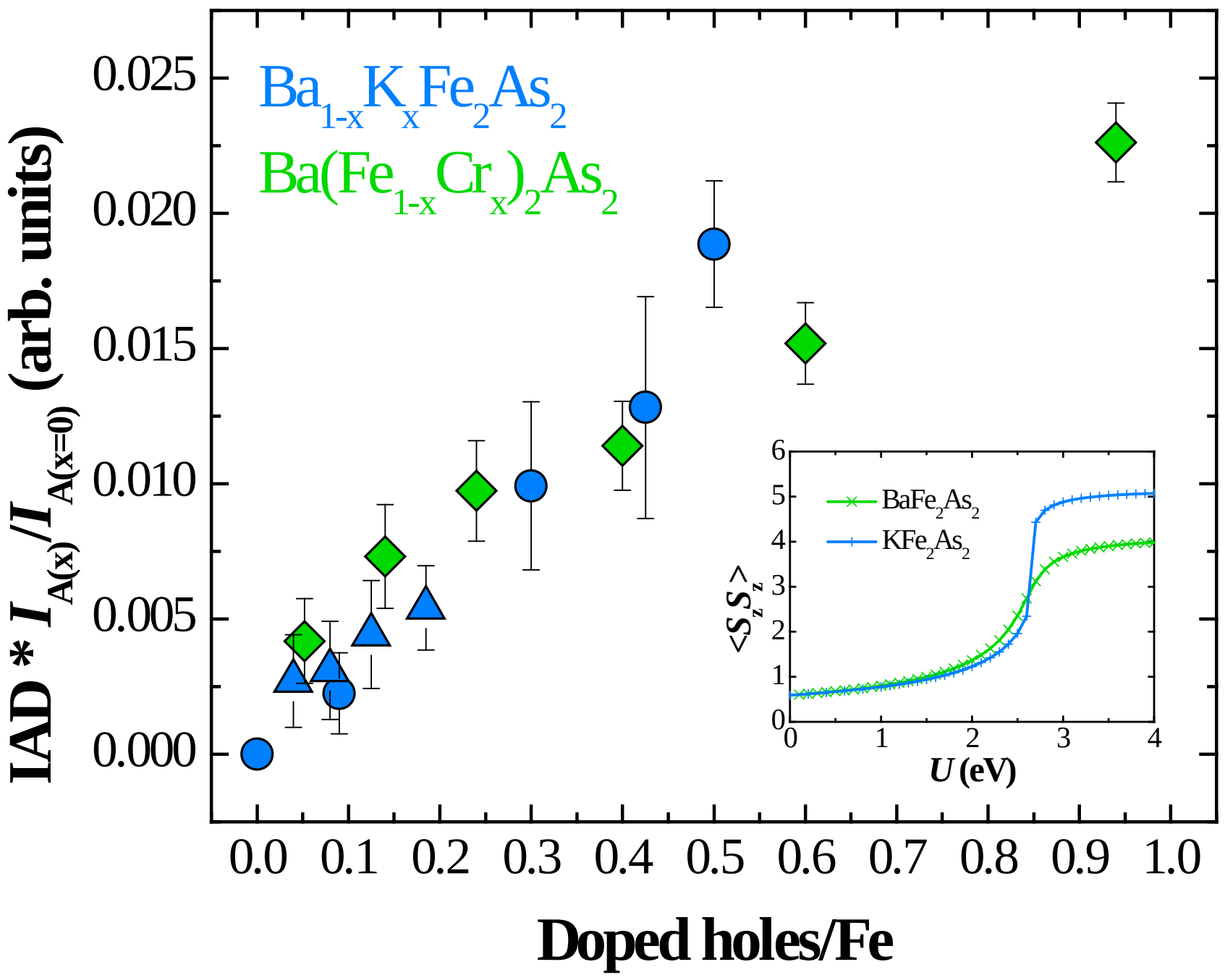

$Ba_{1-x}K_xFe_2As_2$
$Ba(Fe_{1-x}Cr_x)_2As_2$
IAD * $I_{A(x)}/I_{A(x=0)}$ (arb. units)
0.025
0.020
0.015
0.010
0.005
0.000
0.0 0.1 0.2 0.3 0.4 0.5 0.6 0.7 0.8 0.9 1.0
Doped holes/Fe
$BaFe_2As_2$
$KFe_2As_2$
$<S_zS_z>$
6
5
4
3
2
1
0
0 1 2 3 4
$U$ (eV)

# Supplemental Material to:
# "Evidences of Mott physics in iron pnictides from x-ray spectroscopy"

S. Lafuerza[1,*], H. Gretarsson[2], F. Hardy[4], T. Wolf[4], C. Meingast[4], G. Giovanneti[5], M. Capone[5], A. S. Sefat[6] Y. –J. Kim[3], P. Glatzel[1,*] and L. de' Medici[1,*]

[1]*ESRF – The European Synchrotron, CS40220, F-38043 Grenoble Cedex 9, France*
[2]*Max-Planck-Institut für Festkörperforschung, Heisenbergstraβe 1, D-70569 Stuttgart, Germany*
[3]*Department of Physics, University of Toronto, 60 St. George St., Toronto, Ontario, M5S 1A7, Canada*
[4]*Karlsruher Institut für Technologie, Institut für Festkörperphysik, 76021 Karlsruhe, Germany*
[5]*CNR-IOM-Democritos National Simulation Centre and International School for Advanced Studies (SISSA), Via Bonomea 265, I-34136, Trieste, Italy*
[6]*Materials Science and Technology Division, Oak Ridge National Laboratory, Oak Ridge, Tennessee 37831-6114, USA*
*E-mail: sara.lafuerza@esrf.fr, demedici@esrf.fr, pieter.glatzel@esrf.fr

In this supplement we provide additional information about the results of Fe $K\beta$ valence-to-core (VTC) XES measurements, simulations of the HERFD-XANES spectra across the Fe *K* edge and Slave-spin mean-field calculations of the local moment.

## Fe K$\beta$ VTC XES spectra

To complete the $K\beta$ XES study, we also measured the VTC $K\beta$ lines, also referred to as satellite emission lines. While the CTC $K\beta$ lines are sensitive to the local spin magnetic moment, VTC $K\beta$ lines mainly probe the occupied metal *p*-density of states (DOS) up to 25 eV below the Fermi energy which is strongly mixed with ligand *s* and *p* orbitals [1]. These emission lines correspond to transitions from valence orbitals with metal *p* character and the spectrum can be divided into the $K\beta_{2,5}$ (main peak) and $K\beta''$ (feature at lower emission energies) regions. Figures S1(a) and S1(b) show the VTC $K\beta$ spectra separately for the $Ba_{1-x}K_xFe_2As_2$ and $Ba(Fe_{1-x}Cr_x)_2As_2$ series after background removal. The VTC spectra were further treated by performing a background subtraction in order to remove the contribution from the CTC main line tail. For that, the background was modelled using several Voigt functions that were fitted to several data points above (7120-7130 eV) and below (7060-7085 eV) the VTC features following the procedure described in [1].

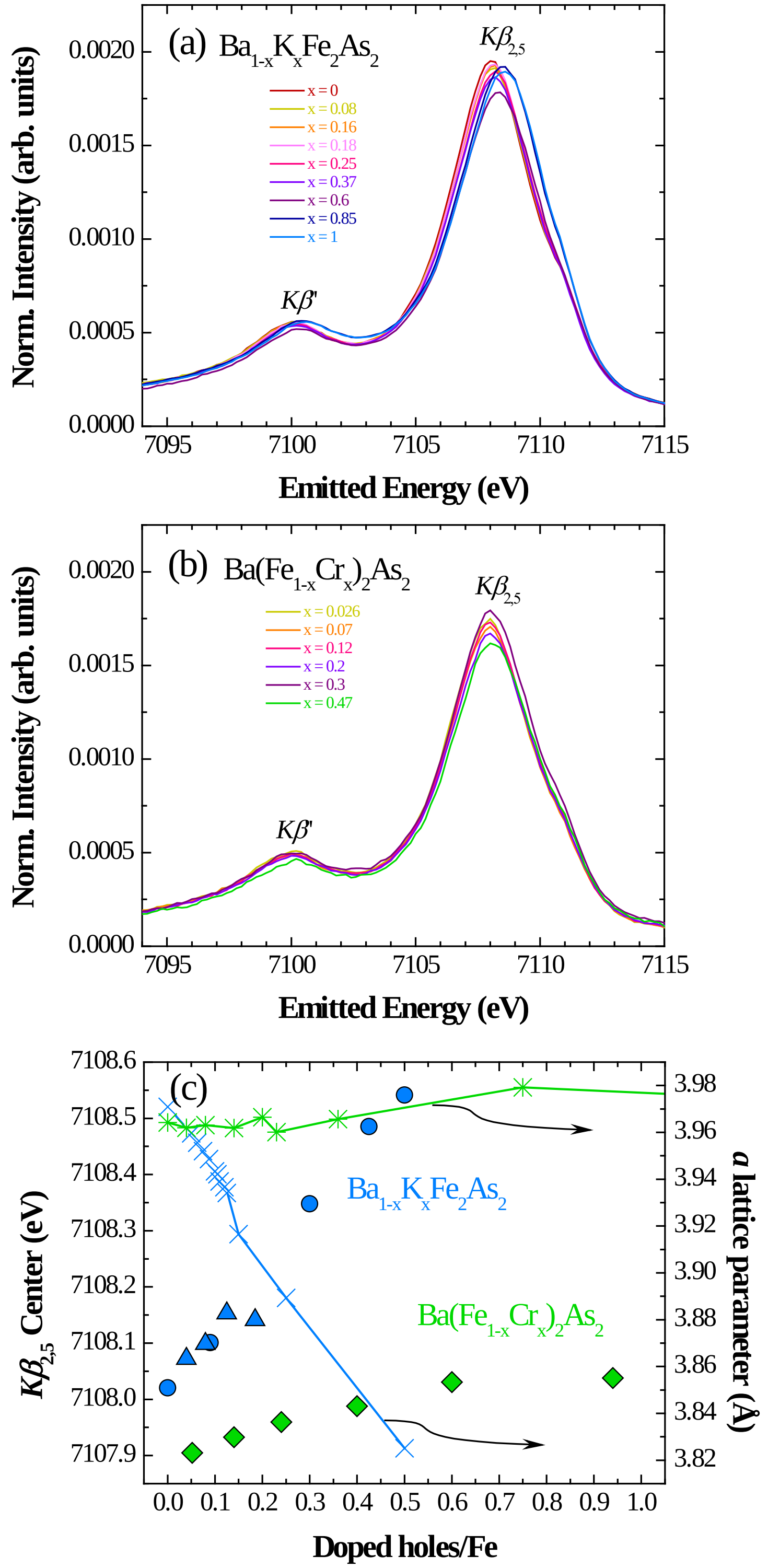

**Figure S1.** Normalized Fe $K\beta$ VTC XES spectra of the (a) $Ba_{1-x}K_xFe_2As_2$ ($0 \leq x \leq 1$) and (b) $Ba(Fe_{1-x}Cr_x)_2As_2$ ($0.026 \leq x \leq 0.47$) series after background removal. (c) Evolution of the $K\beta_{2,5}$ peak center position in energy (left axis) as a function of doped holes/Fe. Circles and triangles differentiate the K-doped samples from sets 1 and 2 respectively. On the right axis the data of the tetragonal *a* lattice parameter from refs. [2] and [3] is plotted for a direct comparison.

Upong K-doping the intensity of the features decreases until x = 0.6, from where a recovery in the intensity is observed coupled with a shift towards higher energies. The intensity decreases as well upon Cr-doping but no major shift is observed. The full VTC spectra were fitted using four Lorentzian functions to analyze the $K\beta_{2,5}$ and $K\beta''$ peaks. The evolution of the $K\beta_{2,5}$ peak center position in energy is shown in Figure S1(c) as a function of holes/Fe. A similar trend is observed for the $K\beta''$ peak center (not plotted). We have included in the same figure the evolution of the in-plane *a* lattice parameter taken from diffraction experiments results reported in refs. [2] and [3] for $Ba_{1-x}K_xFe_2As_2$ and $Ba(Fe_{1-x}Cr_x)_2As_2$ respectively. While little variation is observed with Cr-doping, the lattice parameter shrinks upon K-doping with a consequent shortening of the Fe-As and Fe-Fe distances. We observe a shift to higher energies of the $K\beta_{2,5}$ line with increasing K-content (about 0.5 eV shift between x = 0 and x = 1). This means that shorter bond lengths increase the Fe-ligand orbital mixing and push the occupied *p*-DOS to higher energies. Conversely, $Fe^{2+}$ substitution by $Cr^{2+}$ directly in-plane barely affects the lattice constant compared with K-doping. Accordingly, the $K\beta_{2,5}$ line does not shift notably with Cr-doping.

**Simulations of the HERFD-XANES spectra across the Fe *K* edge**

With the aim of gaining more insights into the origin of the different spectral features and their evolution with doping we performed simulations in the framework of the multiple scattering (MS) theory using the FDMNES code [4]. That is, density functional theory calculations in the local density approximation (DFT-LDA) and one-electron transitions. The tetragonal crystal structures (space group *I4/mmm*) of $BaFe_2As_2$ [5] and $KFe_2As_2$ [2] were used as input in the simulations, carried out in real space with a muffin-tin approximation for the potential. For the exchange–correlation part, the real Hedin, Lundqvist and Von Barth potential was used. The spectra were convoluted using an appropiate Lorentzian function. To compare with the experiment, the computed energy values in the calculations were aligned with the experimental energy scale with the same rigid shift so that the first peak coincides for the $KFe_2As_2$ compound. The cluster size around the central absorber Fe atom that is considered in the calculations is progressively increased in order to monitor the effect of each neighbouring atom shell on the spectral features.

Figure S2(a) shows the calculated spectra for $BaFe_2As_2$ and $KFe_2As_2$ obtained for a cluster radius of 6.6 Å (about 50 atoms) which corresponds to the size beyond which only minor spectral changes are observed. A third calculation is included in which the atomic positions given by the $BaFe_2As_2$ structure were used but with 50% of the Fe atoms randomly replaced by Cr in order to mimic the x = 0.47 Cr-doping effect. The calculations nicely reproduce the experimental spectral line shape. The *A* peak appears in the calculations when the four As nearest neighbours conforming the $FeAs_4$ tetrahedra (at 2.40 Å and 2.38 Å for the $BaFe_2As_2$ and $KFe_2As_2$ structures respectively) are included in the cluster, corroborating that it arises from the mixing between the Fe 3*d* and As 4*p* orbitals. Substitution of Fe by Cr slightly reduces this 3*d*-4*p* orbital mixing as illustrated in the simulation for Fe/Cr in 1:1 proportion, where a weaker *A* peak is observed in accordance with the experimental data. Moreover, the *A* peak was found to be of predominantly dipolar character since when adding quadrupolar transitions in the calculations only a tiny intensity increase (lower than 5 %) was obtained. For $KFe_2As_2$, the *B* feature appears when the first four K neighbours at 3.94 Å are included. Similarly, for $BaFe_2As_2$ the *C* feature is seen once the first four Ba neighbours at 3.81 Å are considered. The *D* - *F* spectral features at higher incident energy result from contributions of neighbouring atoms beyond the first coordination shells and markedly differ between $BaFe_2As_2$ and $KFe_2As_2$. Finally, in order to check the K-doping effect in the simulations, Ba(K) atoms were randomly replaced by K(Ba) in the cluster for 50 and 100% substitution rates while maintaining the crystal structure of $BaFe_2As_2$($KFe_2As_2$). As shown in Figures S2(b) and S2(c), we find that these calculations reproduce the main changes in the experimental spectra upon K-doping. This result indicates that, in addition to the structural changes, the presence of a particular electronic configuration in the tetragonal structure is also responsible for the enhanced Fe 3*d*-As 4*p* orbital mixing.

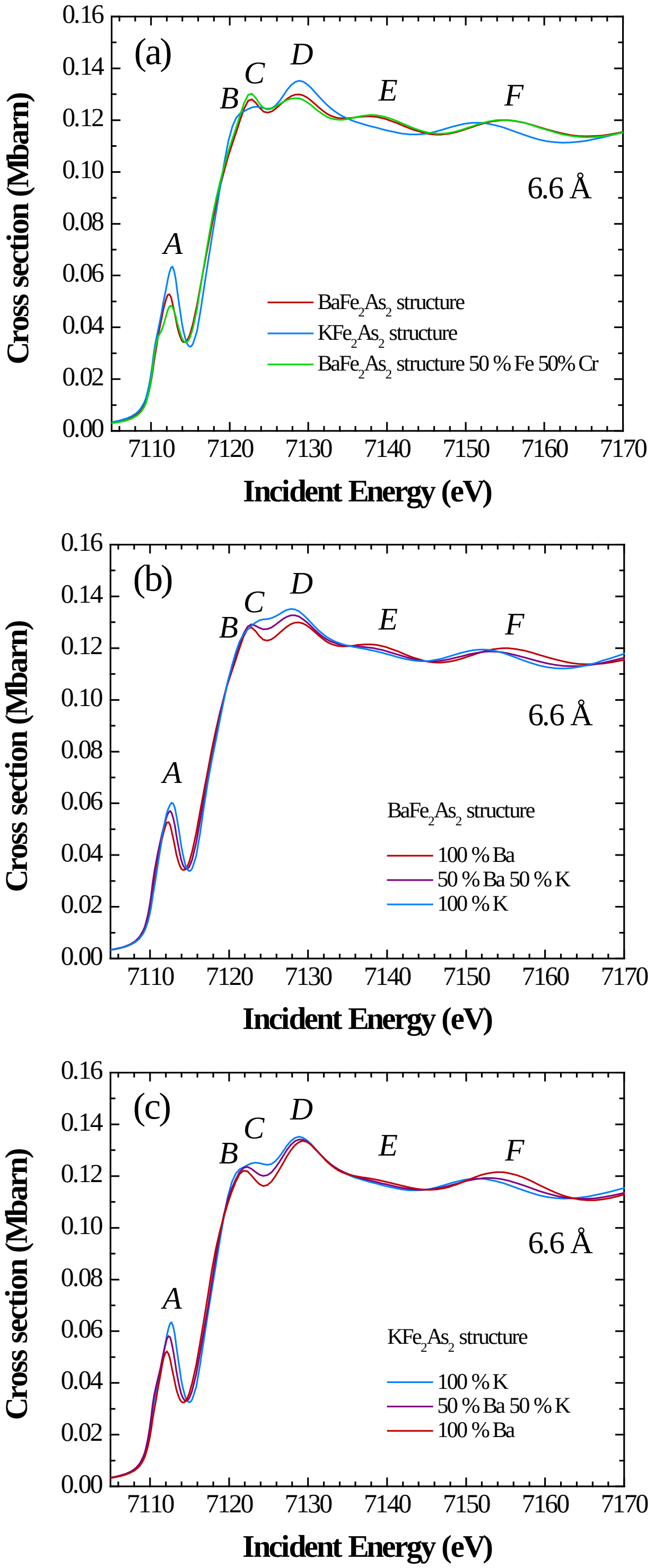

**Figure S2.** (a) Calculated XANES spectra for a cluster size of 6.6 Å around Fe (about 50 atoms) using the tetragonal crystal structures of $BaFe_2As_2$ and $KFe_2As_2$. To simulate the effects of Cr-doping Fe atoms were randomly replaced by Cr for 50 % substitution rate while keeping the atomic positions of the $BaFe_2As_2$ structure. (b) and (c) panels show the effect of replacing Ba and K atoms in the clusters while maintaining the atomic positions of the $BaFe_2As_2$ and $KFe_2As_2$ crystal structures respectively for 50 and 100 % substitution rates.

**Slave-spin mean-field theoretical calculations**

Material-specific many-body calculations of the local moment as a function of the electron-interaction strength reported in the inset of Fig. 3 are performed using Density Functional Theory with the Generalized Gradient Approximation for the exchange-correlation potential according to the Perdew-Burke-Ernzerhof recipe as implemented in Quantum Espresso [6]. The Wannier90 code [7] is used to find the basis of maximally localized Wannier orbitals (10 orbitals, 5 orbitals/Fe) describing the conduction bands. The tight-binding Hamiltonian for the conduction electrons in this basis is supplemented by an explicit local interaction term *à la Hubbard* reading:

$$H_{int} = U \sum_{i,m} n^d_{im\uparrow} n^d_{im\downarrow} + (U-2J) \sum_{i,m>m',\sigma} n^d_{im\sigma} n^d_{im'\bar{\sigma}} + (U-3J) \sum_{i,m>m',\sigma} n^d_{im\sigma} n^d_{im\sigma}$$

where $n^d_{im\sigma}$ is the number operator for electrons in orbital m of site i with spin $\sigma$, and $\bar{\sigma}$ is the opposite of $\sigma$. This multi-orbital Hubbard model is solved within slave-spin mean-field theory [8,9] (see supplemental material of Ref. [1] in the main text for further details).

The quantity reported in Fig. 3, $< S_z S_z >$, is the equal-time correlation function between the z-component of the total spin on the Wannier orbitals. This is easily related the total spin "on site" in the uncorrelated limit, since

$$< S^2 >=< S_x S_x > + < S_y S_y > + < S_z S_z >=3< S_z S_z >,$$

because of rotational invariance.

The interaction term in the density-density approximation used here breaks this rotational invariance so that the previous equality does not hold at finite *U*. There, it is easy to understand that a larger component of the correlation function is found in the z direction, thanks to the fact that the interaction couples the spins in this direction. Nevertheless the correlation function remains a monotonous function of *S*, and a saturated value of the correlation function represents a saturated value of the total local spin to its maximum value.